\documentclass[sigconf]{acmart}

\AtBeginDocument{%
  }

\setcopyright{acmlicensed}
\copyrightyear{2018}
\acmYear{2018}
\acmDOI{XXXXXXX.XXXXXXX}
\acmConference[Conference acronym 'XX]{Make sure to enter the correct
  conference title from your rights confirmation email}{June 03--05,
  2018}{Woodstock, NY}
\acmISBN{978-1-4503-XXXX-X/2018/06}

\usepackage{colortbl}
\usepackage{algorithm}
\usepackage{algorithmic}
\usepackage{amsfonts}
\usepackage{amsmath}
\usepackage{amssymb}
\usepackage{balance}
\usepackage{bbding}
\usepackage{bibentry}
\usepackage{booktabs} % for professional tables
\usepackage{caption}
\usepackage{cases}
\usepackage{color}
\usepackage{enumitem}
\usepackage{epstopdf}
\usepackage{epsfig}
\usepackage{etoolbox}
\usepackage{float}
\usepackage{graphicx}
\usepackage{mathtools}
\usepackage{microtype}
\usepackage{multicol}
\usepackage{multirow}
\usepackage{pifont}
\usepackage{rotating}
\usepackage{subfigure}
\usepackage{xcolor}
\usepackage{enumerate}

\usepackage{bm}

\begin{document}

% \title{MiniRec: Small Data Subsets for Efficient RL-based Large Language Model Recommendation}

\title{MiniRec: Data-Efficient Reinforcement Learning for LLM-based Recommendation}


\author{Lin Wang}
% \authornote{Equal contributions.}
\affiliation{
	\institution{The Hong Kong Polytechnic University}
  \city{Hong Kong}
  \country{China}
}
\email{comp-lin.wang@connect.polyu.hk}
\author{Yang Zhang}
\authornote{Corresponding author.}
\affiliation{
	\institution{National University of Singapore}
  \country{Singapore}
}
\email{zyang1580@gmail.com}

\author{Jingfan Chen}
\affiliation{
	\institution{The Hong Kong Polytechnic University}
  \city{Hong Kong}
  \country{China}
}
\email{jingfan.chen@connect.polyu.hk}
\author{Xiaoyan Zhao}
\affiliation{
	\institution{The Chinese University of Hong Kong}
  \city{Hong Kong}
  \country{China}
}
\email{	xzhao@se.cuhk.edu.hk}

\author{Fengbin Zhu}
\affiliation{
	\institution{National University of Singapore}
  \country{Singapore}
}
\email{fengbin@nus.edu.sg}

\author{Qing Li}
\affiliation{
	\institution{The Hong Kong Polytechnic University}
  \city{Hong Kong}
  \country{China}
}
\email{csqli@comp.polyu.edu.hk}
\author{Tat-Seng Chua}
\affiliation{
	\institution{National University of Singapore}
  \country{Singapore}
}
\email{dcscts@nus.edu.sg}


\renewcommand{\shortauthors}{Lin Wang et al.}


\begin{abstract}
  % The integration of reinforcement learning (RL) into large language models (LLMs) has recently opened new opportunities for recommender systems, 
  % enabling more adaptive and generative user preference modeling. 
  % However, RL-based LLM recommendation remains hindered by severe data inefficiency and high training cost, 
  % as existing methods rely on costly reward computation and suffer from unreliable data pruning strategies. 
  % In this work, we systematically analyze the challenges of subset selection for RL-based LLM recommendation, 
  % identifying two key obstacles: (i) ineffective estimation of a sample's true optimization contribution and 
  % (ii) the misalignment between embedding-space representativeness and semantic training value. 
  % To overcome these challenges, we propose MiniRec, a data pruning framework that constructs a unified value function integrating learnability, representativeness, 
  % and diversity for iterative subset selection. Furthermore, MiniRec employs curriculum learning to stabilize training and enhance generalization by ordering subsets from easier to harder samples. 
  % Extensive experiments on multiple real-world recommendation datasets demonstrate that 
  % MiniRec achieves comparable or superior performance to full-data training with only a fraction of the data 
  % and substantially reduced cost. 
  % This work sheds light on efficient data usage in RL-based LLM recommendation and provides a principled approach 
  % toward practical and scalable deployment.
The integration of reinforcement learning (RL) into large language models (LLMs) has opened new opportunities for recommender systems by eliciting reasoning and improving user preference modeling. However, RL-based LLM recommendation faces significant efficiency challenges, making full-data training costly. Existing data selection methods define sample value based on learnability or representativeness, yet their loss/gradient-driven or dataset coverage-driven criteria often misalign with RL learning dynamics, resulting in suboptimal performance.

To address this, we propose MiniRec, a data selection framework tailored for RL-based LLM recommendation. MiniRec evaluates sample learnability using key RL signals—rewards—pruning samples that are too easy (too high reward) or too difficult (consistently low reward). It assesses representativeness by aligning sample gradients with the approximated  ``ideal'' global RL optimization trajectory, selecting samples that mainly drive model updates, and it also enforces diversity to reduce redundancy. Combined with a curriculum learning strategy from easy to hard samples, MiniRec significantly reduces training cost while largely preserving performance. Extensive experiments demonstrate MiniRec’s effectiveness, highlighting the importance of reward-aligned, trajectory-informed data selection in RL-based LLM recommendation. Our code are available at \href{https://anonymous.4open.science/r/MiniRec-4323/README.md}{https://anonymous.4open.science/r/MiniRec-4323/README.md}

% \textbf{Relevance}:This work advances Web-scale intelligent recommendation by improving the efficiency of RL-based LLM recommenders widely used in online platforms. Our proposed MiniRec framework enables scalable and reasoning-aware recommendation learning through efficient data selection aligned with reinforcement learning optimization. The approach directly contributes to Web intelligence and recommendation research by reducing training costs while maintaining high-quality personalized recommendations.

\end{abstract}




\maketitle

\section{Introduction}
The rapid advancement of large language models (LLMs) is reshaping recommender systems into a new paradigm~\cite{tallrec,llmrecsurvey,collm,acharya2023llm,wei2024llmrec}.
Recent studies show that incorporating reinforcement learning (RL) to elicit stronger reasoning in LLMs can further enhance their ability to infer user preferences and thereby improve overall recommendation performance~\cite{latentR3,you2025text,Reason2rec,wang2025llm}.
However, RL-based LLM recommendation faces a significant efficiency bottleneck, as advanced optimization algorithms such as GRPO~\cite{grpo, deng2025exploring} often require generating multiple trajectories per sample~\cite{Reason2rec}, leading to high GPU memory consumption and prolonged training time.
A natural solution to reduce computational cost is to train on a subset of the data~\cite{DEALRec}. Yet, the RL model's performance heavily depends on the quality of the selected samples.
This raises a key question: how can we effectively select training subsets while maintaining strong model performance?   

% ***********************************************

Existing works~\cite{lin2024data,wu2023leveraging,zheng2022coverage,paul2021deep} on data selection (or data pruning) in recommender systems can be divided into two categories based on how they identify the most valuable samples:
1) Learning signal–driven approaches, which select “hard” or “forgotten” samples based on gradient norms or loss signals; and
2) Similarity/infor\-mation-driven approaches, which select samples to ensure coverage of the data through clustering or submodular optimization.
Learning signal–driven methods focus on identifying the learnability of each sample and its contribution to the learning process\cite{wang2023gradient,mei2025goracs}. In contrast, similarity-driven methods emphasize sample representativeness, aiming to select subsets that can stand in for the full dataset and achieve model optimization performance close to training on all data~\cite{albalak2024survey}. 
Both lines of work have demonstrated significant success in conventional recommendation scenarios.

\begin{figure}[t]
    \centering
    \includegraphics[width=0.47\textwidth]{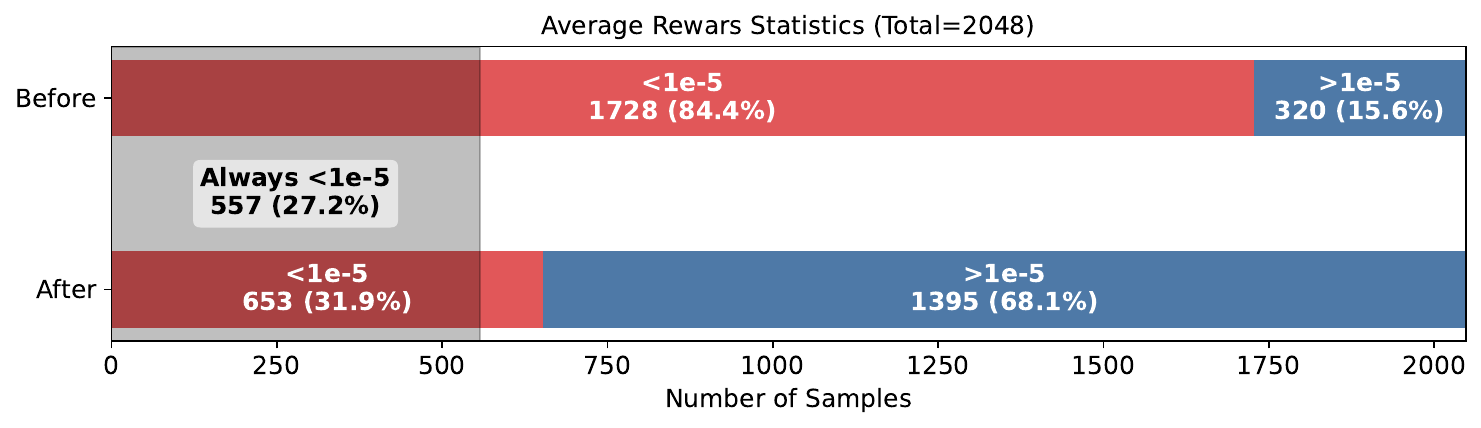}
    \caption{Comparison of average rewards distributions before and after training. 
    Each horizontal bar shows the proportion of samples with very low average rewards (red) and normal rewards (blue). 
    The shaded band highlights the subset of 557 samples (27.2\%) that 
    stayed at very low rewards both before and after training, 
    indicating that more than one quarter of the dataset consistently failed to achieve any optimization contribution throughout training.}
    \label{fig:hit}
\end{figure}

However, we identify two fundamental challenges for these data selections for application in RL-based LLM recommendation: 
\begin{itemize}[leftmargin=*]
    \item Loss and gradients fail to measure sample learnability. In advanced RL methods such as GRPO, the loss and corresponding gradients depend on the relative reward comparisons among trajectories for each sample. This relative comparison can introduce noise into the assessment of sample learnability, causing samples with consistently low rewards to contribute disproportionately to the gradient norm and loss. As shown in Figure~\ref{fig:hit}, over one-quarter of samples consistently exhibit very low rewards during the training process, yet correspond to high loss and gradient norms. Clearly, paying too much attention to these samples does not meaningfully improve learning.

    \item Dataset coverage does not guarantee representativeness for RL optimization. Unlike conventional recommendations, RL-based LLM recommendation is typically designed to teach the model to first reason and then generate outputs, \textit{i.e.,} $ \textit{input} \rightarrow \textit{reason} \rightarrow \textit{output} $, rather than directly $ \textit{input} \rightarrow \textit{output} $. Simply ensuring coverage of $ (\textit{input},\textit{output})$ data pairs potentially fails to capture the samples most representative for learning reasoning. As shown in Table~\ref{tab:kmeans}, we empirically find that the classical coverage-based selection method can even underperform random sampling.
    
    % Simply focusing on coverages fails to measure the sample representativeness for RL optimization. Different from conventional recommendations, the RL-based recommendation needs the model to first learn to reason and then generate the output, \textit{i.e.,} $ \textit{input} \rightarrow \textit{reason} \rightarrow \textit{output} $, instead of directly $\textit{input} \rightarrow \textit{output}$. In this case, ensuring the coverage of the dataset, \textit{i.e.,} the distribution of $\textit{input} \rightarrow \textit{output}$, does not necessarily capture the representative part of $ \textit{input} \rightarrow \textit{reason} \rightarrow \textit{output}$. As shown in Table~\ref{}, such a method (clustering-based) could even underperform the random selection method.  
\end{itemize}

% The challenges highlight the need for a new paradigm that can 
To address these challenges, it is essential to reliably estimate sample learnability and representativeness in alignment with the RL learning.
For learnability, since the reward directly reflects the model’s learning state, it is natural to leverage it for identifying the samples' learnability. 
For representativeness, the key lies beyond the 
$\textit{input}\rightarrow \textit{output}$ relation and instead involves considering the 
$\textit{input}\rightarrow \textit{reason} \rightarrow \textit{output}$ relation. 
However, because the reliable “reason” is difficult to obtain, defining representativeness from a data-coverage perspective becomes infeasible. 
Given that the optimization trajectory of an RL model—from the initial to the final updated model—encodes the learning of the 
$\textit{input} \rightarrow \textit{reason} \rightarrow \textit{output}$, we could define samples whose gradients are more aligned with this trajectory as representative for reasoning learning. By approximating this ideal optimization direction, the approach removes the need to explicitly access the detailed “reason” data.

Based on these insights, we introduce \textbf{\textit{MiniRec}}, a data selection framework tailored for RL-based LLM recommendation. Specifically, MiniRec leverages a proxy model to efficiently estimate sample rewards for learnability definition, pruning samples with low learning value—those that are too easy or too hard to learn, \textit{i.e.,} those with too high or always too low samples. Simultaneously, it uses global second-order gradients to approximate the ideal model optimization direction, measuring sample representativeness by the alignment between each sample’s gradients and this ideal direction, and then selecting samples that drive the main model updates. 
In addition to learnability and representativeness, MiniRec also incorporates a diversity control mechanism to dynamically adjust the scores of remaining samples, preventing the repeated selection of important but highly similar samples. All of these components together constitute the sample value measurement and selection mechanism in MiniRec.

\begin{table}[t]
    \caption{Training results on \texttt{Gemma-2-2b-it} with 1,024 samples. 
    Random selection outperforms embedding-based K-means, 
    suggesting that embedding-space representativeness does not align with  training value. 
    Metrics: NDCG (N@k) and Hit Ratio (HR@k).}
    \label{tab:kmeans}
    \setlength{\tabcolsep}{3pt} % 默认是 6pt，可调小到 3pt-4pt
    \begin{tabular}{ccccccc}
    \hline
    \multirow{2}{*}{Method} & \multicolumn{6}{c}{CDs and Vinyl (1024)}              \\
                            & N@5 & N@10 & N@20 & HR@5   & HR@10  & HR@20  \\ \hline
    Random                  & 0.0297 & 0.0361  & 0.0419  & 0.0459 & 0.0657 & 0.0888 \\
    Kmeans                  & 0.0269 & 0.0349  & 0.0401  & 0.0445 & 0.0654 & 0.0875 \\ \hline
    \end{tabular}
\end{table}

Beyond data selection, MiniRec further employs curriculum learning to maximize the utility of the selected data. 
The selected subset is partitioned into batches, and training proceeds progressively from easier to harder samples. 
This curriculum schedule stabilizes early-stage optimization by preventing early exposure to overly difficult samples, 
while gradually exploiting the richer reward signals embedded in harder samples. With the unified value function, 
curriculum learning improves optimization stability and generalization.

Our contributions are threefold:
\begin{itemize}[leftmargin=*]
\item We systematically analyze the challenges of data pruning in RL-based LLM recommendation and propose principles to evaluate sample value.
\item We develop \textbf{MiniRec}, a framework that integrates learnability, representativeness, and diversity into data selection while accounting for RL-specific characteristics, and further incorporates curriculum learning to enhance training efficiency.
\item Extensive experiments on multiple real-world recommendation datasets show that MiniRec achieves comparable or superior performance while significantly reducing training cost compared to full-data training.
\end{itemize}

\section{Related Works}

\subsection{LLM based Recommendation}

Recent progress in large language models (LLMs) has profoundly reshaped recommender system research. 
Early neural sequential recommenders, such as SASRec~\cite{kang2018self} and BERT4Rec~\cite{sun2019bert4rec}, established powerful contextual encoding foundations for user behavior modeling. 
With the rise of LLMs, studies have explored extending these capabilities to more interactive and explainable paradigms. 
Chat-Rec~\cite{gao2023chat} introduces conversational augmentation of LLMs for interactive feedback understanding, while LLMRec~\cite{liu2023llmrec} systematically benchmarks LLMs across multiple recommendation tasks. 
To capture richer semantics, RLMRec~\cite{ren2024representation} and P4R~\cite{chen2024prompting} employ LLMs to generate high-quality user/item textual profiles and align the semantic and collaborative spaces through cross-view representation learning. 
Further, Reason4Rec~\cite{fang2025reason4rec} and R$^{2}$ec~\cite{you2025text} integrate deliberate reasoning and user preference alignment, pushing LLM-based recommenders toward cognitively enhanced understanding. 
Beyond prompting and SFT paradigms, Rec-R1~\cite{lin2025recr1} and RLHF-based tuning~\cite{yang2025rlhf} leverage reinforcement learning to directly optimize LLMs with feedback signals. 
To reduce training cost, data-efficient methods such as TF-DCon~\cite{wu2024tfdcon} and data-pruned fine-tuning~\cite{lin2024data} explore LLM-aided dataset condensation and few-shot adaption. 
Finally, TransRec~\cite{lin2024bridging} proposes a transition paradigm to unify language and item identifiers, bridging the semantic gap between LLMs and traditional recommender spaces.

\subsection{Data-Efficient LLM training}

Recent research on large language model (LLM) efficiency has increasingly focused on improving data utilization, instruction tuning, and reasoning efficiency. 
In data selection, MATES~\cite{yu2024mates} introduces a model-aware framework that learns a data influence model to dynamically identify the most beneficial samples during pretraining, reducing computational cost by half. 
Complementary to this, PDS~\cite{gu2025data} formulates data selection as an optimal control problem governed by Pontryagin’s Maximum Principle, providing a theoretically grounded approach for large-scale language model training, while DSIR~\cite{xie2023data} adopts importance resampling in a reduced feature space to efficiently approximate target distributions for large text corpora. 
In instruction tuning, the Flan Collection~\cite{longpre2023flan} demonstrates the crucial role of task balancing and mixed prompting for robust instruction-following capabilities, and LTD Instruction Tuning~\cite{chen2023maybe} further shows that less than 0.5\% of the original dataset can suffice for task-specific adaptation without performance degradation. 
For efficient reasoning, LIMO~\cite{ye2025limo} supports the “Less-is-More” hypothesis that complex reasoning can emerge from a few strategic examples. 
In human preference alignment, difficulty-based selection~\cite{qi2025difficulty} leverages the implicit reward gap in DPO to identify challenging but informative preference pairs, achieving superior alignment using only a fraction of the dataset. 
Together, these works underline a paradigm shift toward data-efficient and model-aware strategies for scaling large language models effectively and sustainably.

\section{Preliminaries}

\subsection{Problem Setup and Notation} 
We denote the full training dataset as $\mathcal{D} = \{x_i\}_{i=1}^N$, 
where $x_i$ refers to an individual sample. 
Our goal is to select a curriculum subset $\mathcal{S} \subseteq \mathcal{D}$ 
with target size $|\mathcal{S}| \ll |\mathcal{D}|$, 
such that training on $\mathcal{S}$ yields similar or better performance 
compared to training on $\mathcal{D}$. 
For each sample $x_i$, 
we compute three types of scores: 
\emph{learnability} $L(x_i)$, 
\emph{representativeness} $R(x_i)$, 
and a diversity term $D(x_i \mid \mathcal{S})$. 
These are combined into a unified value function $V(x_i \mid \mathcal{S})$ 
that guides the selection process. 
Hyperparameters include 
the trade-off factor $\lambda$ controlling the balance between learnability and representativeness, 
and the number of curriculum partitions $K$ used in training.

\subsection{Generalized Reinforced Policy Optimization}

% In this section, we introduce the fundamental concept of reinforcement learning (RL) commonly adopted in large language models (LLMs) for recommendation tasks.

Building upon the proven success of GRPO~\cite{deng2025exploring}, we adopt the Group-based Reinforcement Policy Optimization framework as the foundation of our RL approach. GRPO extends the standard Proximal Policy Optimization (PPO)~\cite{jayant2022model, schulman2017proximal, mnih2016a3c,schulman2017ppo} by incorporating group-based reward normalization, leading to more stable and interpretable policy updates.

Given a reference policy $\pi_{\theta_{\text{old}}}$, GRPO first samples a set of $N$ output trajectories for each input prompt $x$, denoted as ${y_1, y_2, \dots, y_N} \sim \pi_{\theta_{\text{old}}}(y|x)$.
Each trajectory $y_i$ represents a complete reasoning and generation process, encompassing both the model’s intermediate reasoning steps and final output. A scalar reward is then assigned to each trajectory based on a task-specific evaluation function.
\begin{equation}
    r_i = R(x, y_i),
\end{equation}
where $R(\cdot, \cdot)$ quantifies the quality or correctness of the generated output according to predefined criteria.
Unlike conventional PPO, which relies on a learned value baseline, GRPO computes the advantage relative to the group average reward to capture intra-group performance differences:
\begin{equation}\label{eq:advantage}
    A_i = \frac{r_i - \bar{r}}{\sigma},
\end{equation}
where $\bar{r} = \frac{1}{N} \sum_{j=1}^{N} r_j$ denotes the group mean reward, and $\sigma$ is the standard deviation within the group. This normalization mitigates reward variance and reduces bias toward outliers.
The policy parameters $\theta$ are then optimized by maximizing a clipped objective similar to PPO, but using the group-based advantage:
\begin{align}
    & \mathcal{L}_{\text{GRPO}}(\theta) = \\
    & \mathbb{E}_{x, y_i \sim \pi_{\theta_{\text{old}}}}
    \left[
        \min\left(
            \rho_i(\theta) A_i,
            \; \text{clip}\left(
                \rho_i(\theta),
                1 - \epsilon, 1 + \epsilon
            \right) A_i
        \right)
    \right],
\end{align}
where $\rho_i(\theta) = \frac{\pi_\theta(y_i|x)}{\pi_{\theta_{\text{old}}}(y_i|x)}$, and $\epsilon$ is a clipping threshold that restricts the step size of policy updates. This objective encourages improvements for samples that outperform their peers while ensuring stability through bounded updates.

% In summary, GRPO performs policy optimization by comparing each trajectory’s reward against its group-average baseline rather than an absolute reference. This relative formulation enhances robustness and stability, particularly in scenarios with high reward variance.

In summary, GRPO optimizes policy updates by comparing each trajectory’s reward with the group-average baseline, improving robustness under high reward variance. Nevertheless, its reliance on relative rewards may overlook absolute reward signals, potentially limiting guidance toward globally optimal behaviors.

\section{Methodology}

In this section, we present MiniRec, a data selection framework for RL‑based LLM recommendation. It aligns sample learnability and representativeness with RL objectives, integrates diversity control to avoid redundancy, and applies curriculum scheduling to enhance training efficiency and stability.

\subsection{Overview}
\begin{figure*}[t]
  \centering
  \includegraphics[width=0.9\textwidth]{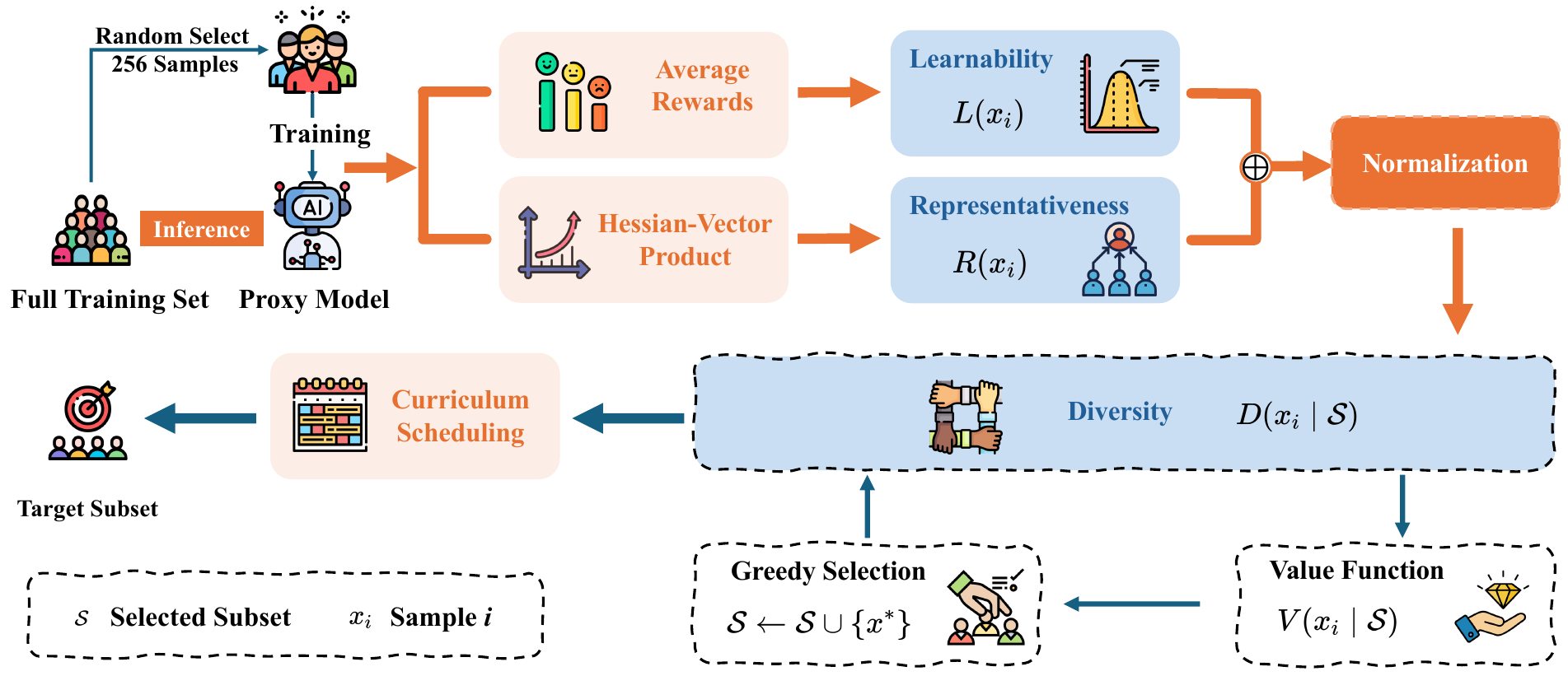}
  \caption{The MiniRec pipeline. 
  We estimate sample-level learnability, representativeness, and diversity 
  from the full training data, normalize these metrics, 
  and integrate them through a value function for greedy subset selection.
  The obtained subset $\mathcal{S}$ is then partitioned for 
  curriculum scheduling to train the model from easy to hard samples.}
  \label{fig:minirec}
\end{figure*}

Figure~\ref{fig:minirec} illustrates the overall pipeline of MiniRec for data selection in RL-based LLM recommendation. As shown, the core idea of MiniRec is to redefine sample learnability and representativeness scores in alignment with RL training, while incorporating a diversity control to prevent overly similar samples and a curriculum learning strategy to better utilize the selected data.

\vspace{+5pt}
\noindent \textbf{Learnability and representativeness.} In this framework, learnability and representativeness scores serve as the core metrics for evaluating data value in selection. To better align them with RL, our key idea is to leverage RL-specific signals that could better reflect RL learning goal in their definition. Specifically,
\begin{itemize}[leftmargin=*]
    \item Learnability score: In RL training, the ultimate goal is to maximize the reward, and thus changes in reward naturally reflect the learnability of samples: if the reward can be easily maximized, the sample is easy to learn; if the reward remains consistently low, the sample is difficult to learn. Accordingly, we define the learnability of a sample based on its reward, estimated via a proxy model. To focus on informative samples, both overly easy and overly hard samples are assigned low scores.

    \item Representativeness score:
In RL-based LLM recommendation, the goal is to teach the model reasoning, capturing the $\textit{input} \rightarrow \textit{reason} \rightarrow \textit{output}$ pattern rather than just $\textit{input} \rightarrow \textit{output}$. To identify representative samples, we focus on those that most effectively support learning this pattern. Specifically, we use the RL model’s optimization trajectory, from initial to final updated models, to capture the learning of $\textit{input} \rightarrow \textit{reason} \rightarrow \textit{output}$. A sample’s representativeness is then measured by how well its gradient aligns with the estimated ideal optimization direction.
    % In contrast, due to the relative advantage computation in the loss and gradients, which could introduce high variance to the loss and gradients, making them no longer a indicator 
\end{itemize}

\noindent \textbf{Diversity control.} Relying solely on learnability and representativeness scores may lead to selecting important but overly similar samples, which adds little additional value. To address this, we dynamically adjust a sample’s value based on its similarity to already selected samples, focusing on choosing novel samples that still have high learnability and representativeness scores.

\vspace{+5pt}
\noindent \textbf{Data usage.} After data selection, rather than using the selected samples blindly, MiniRec applies curriculum learning to maximize their utility—progressively training the model from easier to harder samples.

% In this section, we present the key components of our proposed approach. 
% We first introduce two criteria for evaluating the contribution of each sample to model training, namely \emph{learnability} and \emph{representativeness}. 
% To further enhance the coverage of the selected data, we incorporate \emph{diversity} and formulate a dynamic unified value function that balances all three criteria. 
% Based on this value function, an iterative selection strategy is employed to construct high-quality training subsets, which subsequently guide model optimization.

% For clarity, we introduce the main symbols used throughout the MiniRec framework. 
% The full training dataset is denoted by $\mathcal{D}$, 
% and the candidate pool of not-yet-selected samples is $\mathcal{C}$. 
% The actively selected subset is written as $\mathcal{S}$. 
% Each sample is denoted by $x_i \in \mathcal{D}$, and its representation 
% (e.g., embedding or gradient-based feature) is $f(e_i)$. 
% The per-sample loss is $\ell(x_i; \theta)$ with respect to model parameters $\theta$, 
% and the gradient is $\nabla \ell(x_i)$. 
% These notations will serve as the foundation for later definitions of sample evaluation scores 
% and subset construction strategies.

\subsection{Reward-Based Learnability Estimation}

In RL for LLM-based recommendation, the objective is to optimize the reward. As discussed, loss and gradient are no longer reliable indicators of sample learnability due to the relative advantage computation method, which involves dividing by a variance in Equation~\eqref{eq:advantage}. Instead, we define learnability directly based on the reward that better aligns with the learning goal: if a sample’s reward is easily maximized, it is too easy and contributes little knowledge; if the reward remains consistently low, it is too hard and offers minimal optimization benefit. Therefore, the most informative samples are those of moderate difficulty, with intermediate reward values.

% the supervision signal is typically sparse and derived from \emph{relative} reward comparisons across multiple trajectories. 
% As analyzed in the introduction, this property makes the conventional loss or gradient magnitude unreliable indicators of a sample’s true learnability. 
% Samples that yield large gradients may still correspond to consistently low rewards, offering little optimization benefit, 
% while those consistently achieving high rewards are trivially easy and contribute minimal new information. 
% Hence, the most informative samples for optimization are those of \textit{moderate difficulty}—that is, moderately rewarded samples.

\vspace{+5pt}
\noindent $\bullet$ \textbf{Definition.} For each sample, we define its learnability score based on its reward value, assigning low scores to samples that are either too easy (reward can be easily maximized) or too hard (reward cannot be optimized), and high scores to moderate samples. Formally, for a sample $x_i$, let $\bar{r}(x_i)$ denote the potential optimized reward. We map $\bar{r}(x_i)$ to a continuous learnability score using a Gaussian-shaped function:
\begin{align}\label{eq:gaussian}
L(x_i) = \exp\!\left(-\frac{(\bar{r}(x_i) - \mu)^2}{2\sigma^2}\right),
\end{align}
where $\mu$ represents the optimal “moderate difficulty” point (typically $\mu=0.5$) and $\sigma^2$ controls the tolerance around this target. 

Regarding $\bar{r}(x_i)$, for each sample, it is computed by computing the average reward over different rollout reasoning trajectories. Formally,
\begin{align}\label{eq:reward}
\bar{r}(x_i) = \frac{1}{N}\sum_{n=1}^{N}r_n(x_i),
\end{align}
where $N$ denotes the sampled number of rollout trajectories.

% To operationalize this intuition, MiniRec defines a reward-based learnability score. 
% For each training sample $x_i$, we obtain multiple rollout trajectories and corresponding reward signals. 
% The mean reward over $N$ trajectories is computed as:
% \begin{align}
% \bar{r}(x_i) = \frac{1}{N}\sum_{n=1}^{N}r_n(x_i).
% \end{align}
% where $r_n(x_i)$ denotes the reward obtained from the $n$-th rollout of sample $x_i$, and $N$ is the total number of rollouts.
% We then map the average reward into a continuous learnability score through a Gaussian-shaped function:
% \begin{align}\label{eq:gaussian}
% L(x_i) = \exp\!\left(-\frac{(\bar{r}(x_i) - \mu)^2}{2\sigma^2}\right),
% \end{align}
% where $\mu$ denotes the optimal “moderate difficulty” point (typically $\mu=0.5$) 
% and $\sigma^2$ controls the tolerance around this target. 
% This design ensures that both extremely easy (high-reward) and extremely hard (low-reward) samples are down-weighted, 
% while samples with intermediate reward levels are prioritized as they carry the most valuable optimization signal.

\vspace{+5pt}
\noindent $\bullet$ \textbf{Estimation.} To estimate the learnability score, the key is obtaining the potential optimized reward $\bar{r}(x_i)$. Estimating it using the initial model is not meaningful, as most samples ($>90\%$) receive rewards below 0.1, and using the fully optimized model is infeasible since it is unavailable. To address this, we first train a proxy policy model $\pi_{\theta^{*}}$ on a very small subset (e.g., 256 samples) and use the rewards computed by this proxy to approximate $\bar{r}(x_i)$ for all samples in the training set. These approximated rewards are then applied to Equation~\eqref{eq:gaussian} to compute the scores. Notably, to ensure comparability across samples, the scores are finally normalized to the range $[0,1]$ as,
\begin{align}\label{eq:l_norm}
L_{\text{norm}}(x_i)=
\frac{L(x_i)-\min\limits_{x_j\in\mathcal{D}}L(x_j)}
{\max\limits_{x_j\in\mathcal{D}}L(x_j)-\min\limits_{x_j\in\mathcal{D}}L(x_j)}.
\end{align}
$L_{\text{norm}}(x_i)$ serves as the final learnability score. 
By grounding learnability in reward-level information rather than loss or gradient magnitudes, this design directly aligns with RL training dynamics. For $\bm{m}$,

\subsection{Ideal Optimization Direction-guided Representativeness Estimation}

Beyond learnability, another important aspect of a sample’s value is its ability to represent the overall optimization of the dataset. In RL-based learning, the goal is to capture $\textit{input} \!\rightarrow\textit{reason}\rightarrow \textit{output}$ relations rather than just $\textit{input}\rightarrow \textit{output}$ relations. Consequently, traditional dataset coverage, which reflects only the $(\textit{input}, \textit{output})$ distribution, no longer accurately measures representativeness. Extending coverage to include the “reason” is infeasible, as such data is typically unavailable. 

\vspace{+5pt}
\noindent $\bullet$\textbf{Definition.} Motivated by the goal of selecting samples that reflect the overall dataset optimization and inspired by the work~\cite{huang2025evaluating}, we define representativeness in the optimization space rather than the dataset space.
The key idea is that the optimization trajectory learned by reinforcement learning implicitly encodes the model’s reasoning patterns— 
thus, samples whose gradients are more aligned with this trajectory are more representative for reasoning learning. Once the optimization trajectory is obtained, we could obtain the representativeness. This drives our definition of sample representativeness score.
Formally, let $\bm{d}_g$ denotes the ideal model optimization trajectory (or direction), and $\bm{d}_{i}$ denotes the optimization direction produced by sample $x_{i}$. The representativeness score for $x_{i}$ is defined as follows:
\begin{equation}\label{eq:representativeness}
    R(x_i) = \cos\!\big( \bm{d}_{g},\, \bm{d}_{i} \big),
\end{equation}
where a higher value indicates greater importance.

\vspace{+5pt}
\noindent$\bullet$\textbf{Estimation.} It is not easy to estimate the defined representativeness score, as both $\bm{d}_g$ and $\bm{d}_{i}$ are not easy to obtained. Leveraging the proxy model as done by the learnability score estimation becomes no meaning, because the model may cannot represent the global otpmization direction. Inspired by the \textit{influence function}~\cite{koh2017understanding}, we propose used the high-order (second-order) gradients to approximate the $\bm{d}_g$ and $\bm{d}_{i}$, which could better reflects the final optimization directions. 

Specifically, we incorporates second-order information through a {Hessian–Vector Product (HVP)} formulation~\cite{pearlmutter1994hvp, martens2010hessianfree, koh2017influence}.  
For each sample $x_i$, we approximate its $\bm{d}_{i}$ with the HVP as follows:
\begin{align}\label{eq:hvp}
\bm{d}_{i} \approx \text{HVP}(x_i) = H(x_i) \cdot v_i 
&= \nabla^2_{\theta}\ell(x_i; \theta)\, v_i \\
&= \nabla_{\theta}\,(\nabla_{\theta}\ell(x_i; \theta) ^{\top}v_i),
\end{align}
where $H(x_i)$ is the Hessian of the loss with respect to $\theta$, and $v_i = \nabla_{h}\ell(x_i; \theta)$ denotes the gradient of sample $i$, with $\ell(x_i; \theta)$ denoting the loss and $h$ denoting the hidden state of the last token at the final layer for sample $x_i$.  
This operation implicitly captures the local curvature of the optimization landscape along the gradient direction,  
offering a second-order refinement of a sample’s optimization influence—without explicitly forming the full Hessian matrix. For $\bm{d}_{g}$, it is optimized by the global averged HVP, which is formulated as follows:
\begin{align}\label{eq:g_first}
\bm{d}_g = \frac{1}{|\mathcal{D}|}\sum_{x_i \in \mathcal{D}}\text{HVP}(x_i).
\end{align}

The representativeness score is then approximated as follows:
\begin{equation}\label{eq:representativeness}
    R(x_i) \approx \cos\!\big( \bm{d}_{g},\, \bm{d}_{i} \big),
\end{equation}
Similarly, the representativeness scores are normalized into $[0,1]$ following the scaling scheme in Equation~\eqref{eq:l_norm}.  
A higher $R_{\text{norm}}(x_i)$ indicates stronger alignment with the ideal RL optimization direction,  
meaning the sample better encapsulates the reasoning-driven update trajectory.  
By leveraging this gradient-based alignment principle,  
MiniRec effectively selects samples that are truly representative for the reinforcement learning optimization process.

\subsection{Sample Diversity}

While learnability and representativeness help identify the most informative samples aligned with the RL learning, relying solely on these criteria may lead to redundancy—selecting important yet homogeneous samples that contribute limited additional value.
% Samples that are highly learnable and representative often cluster around similar optimization regions,  resulting in subsets dominated by medium-difficulty samples with near-duplicate reasoning trajectories. 
% This redundancy narrows the coverage of the optimization landscape, 
% suppresses exposure to diverse reasoning patterns, and may cause overfitting or convergence stagnation.  
Therefore, another important aspect of MiniRec is to ensure diversity, 
which preserves heterogeneity in the training signals and enables the model to explore a broader reasoning space.

\vspace{+5pt}
\noindent$\bullet$\textbf{Diversity Score Definition}. In contrast to learnability and representativeness—both intrinsic to individual samples—  
diversity is a \emph{relational property} defined with respect to the subset currently being constructed.  
Diversity measures how distinct a candidate sample is from those already selected.  
Assigning a static diversity score independent of the selected set $\mathcal{S}$ would be inappropriate.  
To explicitly encourage non-redundant selection,  
MiniRec formulates diversity as a dynamic marginal value through a min–max strategy\cite{madry2018adversarial, lin2020minmax}, 
enforcing maximal incremental diversity at each selection step.

 Formally, let $f(x_i)$ denote the feature representation of sample $x_i$  
(e.g., the final-layer hidden state or task-specific embedding),  
and $\mathcal{S}$ be the currently selected subset.  The marginal diversity contribution of a candidate $x_i$ is defined as its minimal distance to the subset:
\begin{align}
D(x_i \mid \mathcal{S}) = \min_{x_j \in \mathcal{S}} \,\mathrm{dist}\!\left(f(x_i), f(x_j)\right),
\end{align}
where $\mathrm{dist}(\cdot,\cdot)$ denotes the distance metric, which is instantiated as cosine distance in this paper.
This function yields high values for samples that lie far from the existing subset in the representation space, 
thus directly promoting sample differentiation.
To maintain consistency with other objectives,  
the diversity score is normalized to $[0,1]$, resulting in $D_{\text{norm}}(x_i \mid \mathcal{S})$.

\vspace{+5pt}
\noindent $\bullet$ \textbf{Final Data Selection.} With considering the diversity control, at each step, the overall selection value for a candidate $x_i$ is then defined as:
\begin{align}
V(x_i \mid \mathcal{S})
= D_{\text{norm}}\!\left(x_i \mid \mathcal{S}\right)
\Big[L_{\text{norm}}(x_i) + \lambda \, R_{\text{norm}}(x_i)\Big],
\end{align}
where $L_{\text{norm}}(x_i)$ and $R_{\text{norm}}(x_i)$ denote the normalized learnability 
and representativeness scores, respectively.  
The coefficient $\lambda$ balances the relative emphasis of optimization alignment versus reward sensitivity. The design ensures that each newly selected sample maximizes marginal diversity while emphasizing learnability and representativeness.

Subset construction proceeds iteratively with a greedy selection rule.  
At each step, the sample with the highest overall value is chosen:
\begin{align}
x^{*} = \arg\max_{x_i \in  \mathcal{D}/\mathcal{S}} V(x_i \mid \mathcal{S}),
\end{align}
and added to $\mathcal{S}$ to form new $\mathcal{S} \leftarrow \mathcal{S} \cup \{x^{*}\}$.  
This process continues until the target subset size is reached.  

% The min–max design ensures that each newly selected sample maximizes marginal diversity 
% while reinforcing learnability and representativeness.  
% As a result, the constructed subset maintains broad coverage over the optimization space, 
% preserves diverse reasoning trajectories, 
% and achieves a balanced selection across learning difficulty, optimization alignment, 
% and reasoning diversity—completing MiniRec’s tri-objective sample selection framework.

\subsection{Curriculum Scheduling}

% While the selected subset $\mathcal{S}$ already encapsulates balanced learnability, representativeness, 
% and diversity, directly training on $\mathcal{S}$ in a mixed and unordered manner may not yield 
% optimal optimization dynamics.
When using the selected data to train the model, blindly mixing samples of different difficulty levels can cause interference, leading to unstable gradients and suboptimal convergence, especially for involving learning reasoning. 
To address this, MiniRec introduces a {curriculum scheduling mechanism}~\cite{wang2021survey, park2023acl} 
that structures the learning process from easier to more challenging samples, 
thereby stabilizing optimization while maintaining diversity across stages.  
This design ensures that the model progressively internalizes foundational reasoning patterns 
before being exposed to harder, more heterogeneous cases.  

\begin{figure}[t]
  \centering
  \includegraphics[width=0.45\textwidth]{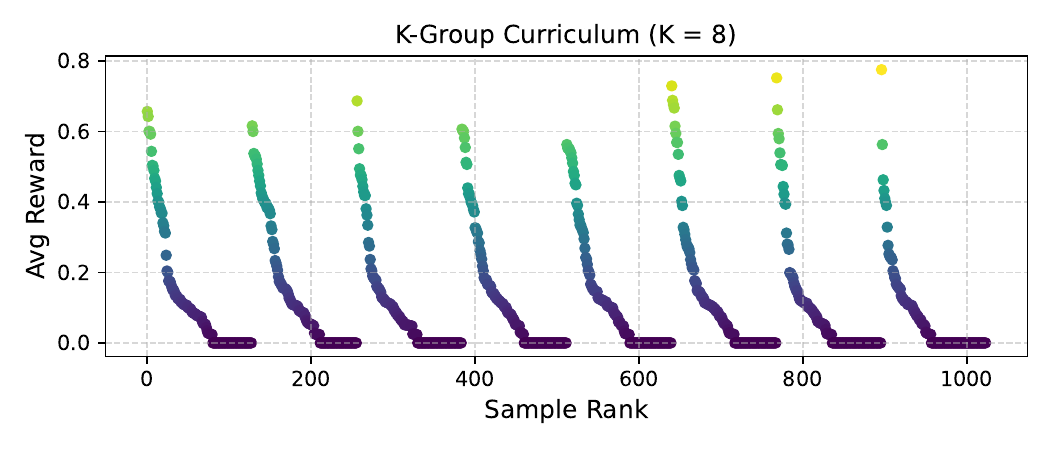}
  \caption{MiniRec’s curriculum scheduling mechanism. 
  % The selected subset $\mathcal{S}$ is randomly partitioned into several groups, and samples within each group are ordered by estimated difficulty (reward values) from easy to hard.
  }
  \label{fig:curriculum}
\end{figure}

To achieve this, as shown in Figure~\ref{fig:curriculum}, the selected subset $\mathcal{S}$ is first randomly partitioned into $K$ disjoint groups:
\begin{align}\label{eq:reward}
\mathcal{S} = \mathcal{S}_1 \cup \mathcal{S}_2 \cup \cdots \cup \mathcal{S}_K, 
\quad \mathcal{S}_i \cap \mathcal{S}_j = \emptyset \;\; (i \neq j).
\end{align}
Within each partition $\mathcal{S}_i$, 
samples are further sorted in descending order of their reward $\bar{r}(x_i)$ defined in Equation~\eqref{eq:reward}, 
ensuring internal progression from easier to more complex instances 
while keeping each group representative of the overall data distribution. 
% \textcolor{blue}{Then, from $\mathcal{S}_{1}$ to $\mathcal{S}_{K}$
% , maintain the order both between and within the groups when sampling minibatch data to train the model.}

Notably, our combination of random partitioning and intra-group ordering is crucial.  
If we globally sort the entire subset $\mathcal{S}$ by difficulty and train the model strictly from easy to hard, 
the training process would become too rigid.  
Because RL-based instruction tuning usually runs for only a few epochs, 
the model would focus heavily on the easiest samples at the beginning, 
quickly fitting their simple patterns and converging to a suboptimal solution.  
Once the model locks into these easy patterns, 
it becomes difficult for later, harder samples to effectively influence learning, 
which can even lead to a decline in performance.  

In contrast, MiniRec’s hybrid scheduling avoids this issue.  
Randomly partitioning the subset keeps each training phase diverse, 
while sorting within each group still provides a clear sense of progression from easier to harder cases.  
This design keeps the curriculum balanced—stable yet flexible—so the model avoids early overfitting, 
benefits from varied learning signals, 
and gradually strengthens its reasoning ability throughout training.

% \section{Implementation Details}
\subsection{Implementation Details}

Algorithm~\ref{alg:minirec} outlines the proposed \emph{Subset Selection with Curriculum Scheduling} framework, which consists of three phases: sample evaluation, subset construction, and curriculum scheduling.
In the first phase, each training sample is evaluated by three criteria—learnability, representativeness, and diversity—to estimate its overall utility.
Next, a diversity-aware greedy selection algorithm iteratively builds a compact yet informative subset, selecting samples that best balance informativeness and complementarity.
Finally, curriculum scheduling organizes the chosen subset into progressive training stages, ordering samples from easy to hard based on estimated difficulty. Starting with simpler instances promotes stable learning and later integration of harder cases improves generalization and robustness.

% Algorithm~\ref{alg:minirec} outlines the Subset Selection with Curriculum Scheduling procedure. The algorithm first computes three complementary metrics for each sample—learnability, representativeness, and diversity—to evaluate its contribution to the subset. A greedy selection strategy iteratively constructs a subset that balances informativeness and coverage under a predefined target size. After selection, the chosen subset is partitioned into multiple curriculum stages based on sample difficulty, enabling the model to be trained progressively from easier to harder samples. This curriculum-driven subset optimization enhances both data efficiency and generalization during training.

\begin{algorithm}[t]
  \caption{Subset Selection with Curriculum Scheduling}
  \label{alg:minirec}
  \begin{algorithmic}[1]
  
  \REQUIRE Full dataset $\mathcal{D}$, base model $M$, 
  target subset size $m$, balance weight $\lambda$, 
  number of curriculum groups $K$.
  
  \vspace{0.5em}
  \STATE \textbf{Learnability:} 
  Estimate rewards $\bar{r}(x_i)$, compute $L(x_i)$ (Eq.~\eqref{eq:gaussian}), 
  and normalize to $L_{\text{norm}}(x_i)$ (Eq.~\eqref{eq:l_norm}).
  
  \STATE \textbf{Representativeness:} 
  Compute representativeness score $R_{\text{norm}}(x_i)$ for each sample.
  
  \STATE \textbf{Diversity-driven Greedy Selection:}
  \STATE Initialize $\mathcal{S} \leftarrow \emptyset$.
  \WHILE{$|\mathcal{S}| <$ target size $m$}
    \FORALL{$x_i \in \mathcal{D} \setminus \mathcal{S}$}
      \STATE Compute diversity score $D_{\text{norm}}(x_i \mid \mathcal{S})$
             as minimum distance to samples in $\mathcal{S}$.
      \STATE Compute overall value:
        \[
        V(x_i \mid \mathcal{S}) = D_{\text{norm}}(x_i \mid \mathcal{S})
        \cdot \big( L_{\text{norm}}(x_i) + \lambda R(x_i) \big).
        \]
    \ENDFOR
    \STATE Select $x^* = \arg\max_{x_i} V(x_i \mid \mathcal{S})$.
    \STATE Update $\mathcal{S} \leftarrow \mathcal{S} \cup \{x^*\}$.
  \ENDWHILE
  
  \STATE \textbf{Curriculum Scheduling:}
  \STATE Randomly partition $\mathcal{S}$ into $K$ groups.
  \STATE Sort $\mathcal{S}_i$ by reward $\bar{r}(x_i)$ (descending).
  \STATE Train $M$ incrementally on $\mathcal{S}_i$ from easy to hard.

  \STATE \textbf{Output:} Final ranked subset  $\mathcal{S}$.
  
  \end{algorithmic}
  \end{algorithm}

\section{Experiments}
\label{sec:experiments}
In this section, we comprehensively evaluate MiniRec through a series of experiments designed to assess its effectiveness, generalization, efficiency, and scalability. We first describe the experimental setup, including datasets and baseline methods, and then present main training results on LLM‑based recommendation. We further analyze MiniRec’s cross‑model generalization, training efficiency, and scalability under varying data budgets. Finally, an ablation study quantifies the contribution of each component, demonstrating that MiniRec’s integrated design is crucial for achieving robust and efficient performance.

% \usepackage{multirow}
\begin{table*}[]
\caption{Performance comparison on the CDs and Vinyl and Instruments datasets with 1,024 training samples. MiniRec consistently outperforms existing data selection baselines (Random, K-means, GraNd, EL2N, DEALRec) across all metrics, approaching full-data performance while using only a subset of data.}
\label{tab:main}
\begin{tabular}{cccccccc}
\toprule
\textbf{Datasets}                                                                            & \textbf{Methods}  & NDCG@5 & NDCG@10 & NDCG@20 & Hit\_ratio@5 & Hit\_ratio@10 & Hit\_ratio@20 \\ \midrule
\multirow{7}{*}{\begin{tabular}[c]{@{}c@{}}\textbf{CDs and Vinyl}\end{tabular}} & \textbf{Random}  & 0.0297 & 0.0361  & 0.0419  & 0.0459       & 0.0657        & 0.0888        \\
 & \textbf{Kmeans}  & 0.0269 & 0.0349  & 0.0401  & 0.0445       & 0.0654        & 0.0875        \\
 & \textbf{GraNd}   & 0.0314 & 0.0384  & 0.0433  & 0.0475       & 0.0667        & 0.0921        \\
 & \textbf{EL2N}    & 0.0323 & 0.0390  & 0.0452  & 0.0509       & 0.0688        & 0.0917        \\
 & \textbf{DEALRec} & 0.0324 & 0.0381  & 0.0429  & 0.0495       & 0.0664        & 0.0902        \\
 & \textbf{MiniRec (Ours)} & 0.0341 & 0.0402  & 0.0464  & 0.0521       & 0.0714        & 0.0960        \\
 & \textbf{Full}    & 0.0345 & 0.0434  & 0.0478  & 0.0521       & 0.0788        & 0.0967        \\ \midrule
\multirow{7}{*}{\begin{tabular}[c]{@{}c@{}}\textbf{Instruments}\end{tabular}}   & \textbf{Random}  & 0.0099 & 0.0114  & 0.0146  & 0.0154       & 0.0201        & 0.0332        \\
 & \textbf{Kmeans}  & 0.0087 & 0.0102  & 0.0122  & 0.0134       & 0.0186        & 0.0321        \\
 & \textbf{GraNd}   & 0.0110 & 0.0159  & 0.0204  & 0.0187       & 0.0301        & 0.0489        \\
 & \textbf{EL2N}    & 0.0123 & 0.0157  & 0.0210  & 0.0196       & 0.0298        & 0.0505        \\
 & \textbf{DEALRec} & 0.0102 & 0.0132  & 0.0177  & 0.0189       & 0.0254        & 0.0436        \\
 & \textbf{MiniRec (Ours)} & 0.0143 & 0.0186  & 0.0223  & 0.0247       & 0.0378        & 0.0545        \\
 & \textbf{Full}    & 0.0161 & 0.0203  & 0.0257  & 0.0264       & 0.0397        & 0.0615        \\ \bottomrule
\end{tabular}
\end{table*}

\subsection{Experimental Setup}

\paragraph{Datasets.}
We adopt two real-world Amazon review datasets, \textbf{CDs \& Vinyl} and \textbf{Instruments}~\cite{you2025text}, which are widely used in recommendation benchmarking. Details please see Table~\ref{tab:dataset_stats}.
Each dataset contains rich user-item interactions and textual reviews, allowing us to assess the quality of selected training subsets across different domains.

\begin{table}[ht]
    \centering
    \caption{Dataset Statistics.}
    \setlength{\tabcolsep}{1pt}
    \begin{tabular}{lcccccc}
    \toprule
    \textbf{Dataset} & \textbf{Users} & \textbf{Items}  & \textbf{Interactions} & \textbf{Train} & \textbf{Val} & \textbf{Test} \\
    \midrule
    CDs and Vinyl        & 7,701  & 12,024  & 13,435 & 10,748 & 1,343 & 1,344 \\
    Musical Instruments  & 15,656 & 10,320  & 34,373 & 27,498 & 3,437 & 3,438 \\
    \bottomrule
    \end{tabular}
    \label{tab:dataset_stats}
\end{table}

% \begin{table}[ht]
%     \centering
%     \caption{Dataset Statistics.}
%     \setlength{\tabcolsep}{5pt}
%     \begin{tabular}{lcc}
%     \toprule
%     \textbf{Statistic} & \textbf{CDs and Vinyl} & \textbf{Musical Instruments} \\
%     \midrule
%     Users          & 7,701  & 15,656 \\
%     Items          & 12,024 & 10,320 \\
%     Density        & 0.023\% & 0.031\% \\
%     Interactions   & 13,435 & 34,373 \\
%     Train          & 10,748 & 27,498 \\
%     Validation     & 1,343  & 3,437 \\
%     Test           & 1,344  & 3,438 \\
%     \bottomrule
%     \end{tabular}
%     \label{tab:dataset_stats}
% \end{table}

\paragraph{Baselines.}
We compare MiniRec with six data selection strategies:
(1) \textbf{Random}: random sampling from the full dataset;
(2) \textbf{KMeans}\cite{ahmed2020k}: clustering-based selection to enhance diversity;
(3) \textbf{GraNd}~\cite{paul2021deep}: gradient-norm–based sample importance;
(4) \textbf{EL2N}~\cite{paul2021deep}: error-based data valuation; and
(5) \textbf{DEALRec}~\cite{DEALRec}: A data pruning method combining influence and effort scores to efficiently select representative samples for few-shot fine-tuning of LLM-based recommendation.
For all methods, the subset size is fixed to $|\mathcal{S}| = 1024$ unless otherwise stated.
\paragraph{Proxy Model.} In general, the proxy model shown in Figure.~\ref{fig:minirec} is kept consistent with the final target model used for training. For example, if the ultimate goal is to fine‑tune the Gemma model, the proxy model is likewise instantiated from Gemma and trained on a small subset of 256 samples. Aligning the proxy and target architectures ensures that the learned data valuation reflects the same optimization dynamics and inductive biases as the final model, thereby improving the reliability and transferability of the selected subset.

% \paragraph{Evaluation Metrics.}
% We evaluate recommendation quality using ranking-based metrics:
% NDCG@5, NDCG@10, NDCG@20, Hit Ratio@5, Hit Ratio@10, and Hit Ratio@20. 
% All results are averaged over three random runs for robustness.

\subsection{Training Performance}
\label{subsec:main-result}

This experiment evaluates MiniRec in training LLM-based recommender models using selected subsets. The objective is to examine whether MiniRec-selected samples can preserve model performance while significantly reducing training data size. We conduct experiments on two Amazon categories—CDs \& Vinyl and Instruments—with subset size $|\mathcal{S}|=1024$. All methods fine-tune the same base LLM Gemma-2-2b-it~\cite{team2024gemma} under identical hyperparameter settings for fairness. Results are reported in Table~\ref{tab:main}.

Across both datasets, MiniRec consistently achieves the best or comparable performance among all baselines. On CDs \& Vinyl, MiniRec attains an NDCG@20 of 0.0464 and a Hit Ratio@20 of 0.0960, closely matching the full-data model. On Instruments, MiniRec also leads with NDCG@20 = 0.0223 and Hit Ratio@20 = 0.0545. These results verify that MiniRec effectively identifies subsets that retain high training informativeness.

In summary, MiniRec’s joint consideration of learnability, representativeness, and diversity enables efficient LLM fine-tuning, delivering robust recommendation quality with minimal data usage.

\subsection{Generalization Across Different LLMs}
\label{subsec:generalization}
\begin{figure*}[t]
\setlength{\abovecaptionskip}{3pt}
\centering
    \centering
    % 指定最大高度或宽度，保证不被裁切
    \includegraphics[width=0.8\linewidth, keepaspectratio]{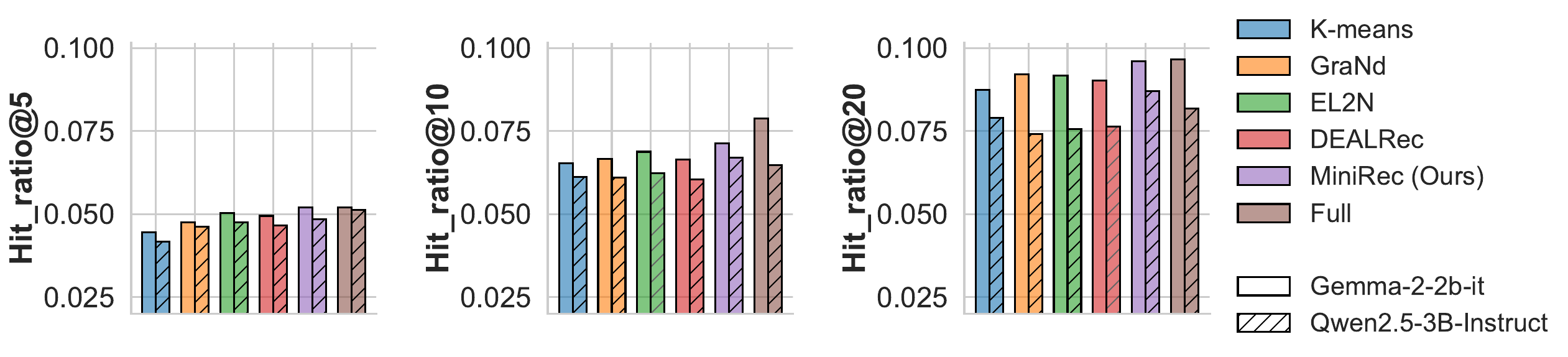}
    \caption{Evaluation of MiniRec’s cross‑model generalization on the CDs \& Vinyl dataset. Subsets selected using one LLM (\texttt{Gemma‑2‑2B‑it}) are transferred to another LLM (\texttt{Qwen2.5‑3B‑Instruct}) for fine‑tuning. MiniRec maintains consistently strong performance across all metrics, demonstrating its ability to capture model‑agnostic and broadly transferable data value.}
    \label{fig:general}
\end{figure*}

% \begin{figure}[h]
%     \centering
%     \includegraphics[width=0.45\textwidth]{figures/generalization.pdf}
%     \caption{Generalization across different LLMs}
%     \label{fig:generalization}
% \end{figure}

To further assess the cross-model generalization ability of MiniRec, we examine whether the subsets selected by one LLM remain effective when transferred to another model. Specifically, we evaluate the effectiveness of subsets $\mathcal{S}$, selected using a proxy model, by fine‑tuning Gemma‑2‑2B‑it~\cite{team2024gemma} on 256 samples, and train separate recommender models based on both Gemma‑2‑2B‑it and Qwen2.5‑3B‑Instruct~\cite{qwen2.5, qwen2} architectures. The subset size is fixed at $|\mathcal{S}| = 1024$, and all models are trained under identical hyperparameter settings to ensure comparability.

Figure~\ref{fig:general} reports the results on the CDs \& Vinyl dataset. Across all evaluation metrics (NDCG@k and Hit Ratio@k, $k \in \{5,10,20\}$), subsets obtained by MiniRec generalize well to unseen LLM architectures. When the subset selected by Gemma‑2‑2B‑it is applied to Qwen2.5‑3B‑Instruct, MiniRec still achieves Hit Ratio@10=0.0670, clearly outperforming other subset selection methods. The consistent leading performance across both models demonstrates that MiniRec-selected samples encapsulate information broadly beneficial to LLM fine-tuning, rather than exploiting model-specific peculiarities.

Overall, these findings confirm that MiniRec’s value function captures intrinsic data value that transfers effectively across different LLM architectures, underscoring its potential as a general-purpose and architecture-agnostic data selection approach for efficient model training.

\subsection{Efficiency Analysis}
\label{subsec:efficiency}
% \usepackage{multirow}
% \begin{table}[]
% \caption{time efficiency}
% \label{tab:time}
% \begin{tabular}{ccccc}
% \hline
% \multirow{2}{*}{Dataset} & \multicolumn{3}{c}{Scale} & \multirow{2}{*}{Full} \\
%                          & 256     & 512    & 1024   &                       \\ \hline
% CDs and Vinyl            & 0.9h    & 1.5h   & 3.3h   & 18.6h                 \\
% Instruments              & 1.5h    & 2.7h   & 5.5h   & 38.5h                 \\ \hline
% \end{tabular}
% \end{table}

% \begin{figure}
%     \centering
%     \includegraphics[width=0.8\linewidth]{figures/time_plot.pdf}
%     \caption{Training time comparison between datasets CDs and Vinyl and Instruments at different data scales (256, 512, 1024, and full).}
%     \label{fig:time}
% \end{figure}

% \begin{figure*}[t]
%     \centering
%     % 左图
%     \begin{minipage}[t]{0.48\textwidth}
%         \centering
%         \includegraphics[width=\linewidth]{figures/Generalization.pdf}
%         \caption{Generalization across different LLMs.}
%         \label{fig:general}
%     \end{minipage}
%     \hfill
%     % 右图
%     \begin{minipage}[t]{0.48\textwidth}
%         \centering
%         \includegraphics[width=\linewidth]{figures/time_plot.pdf}
%         \caption{Training time comparison on the CDs and Vinyl dataset at different data scales (256, 512, 1024, and full).}
%         \label{fig:time}
%     \end{minipage}
% \end{figure*}

\begin{figure}
    \centering
    \includegraphics[width=0.65\linewidth, keepaspectratio]{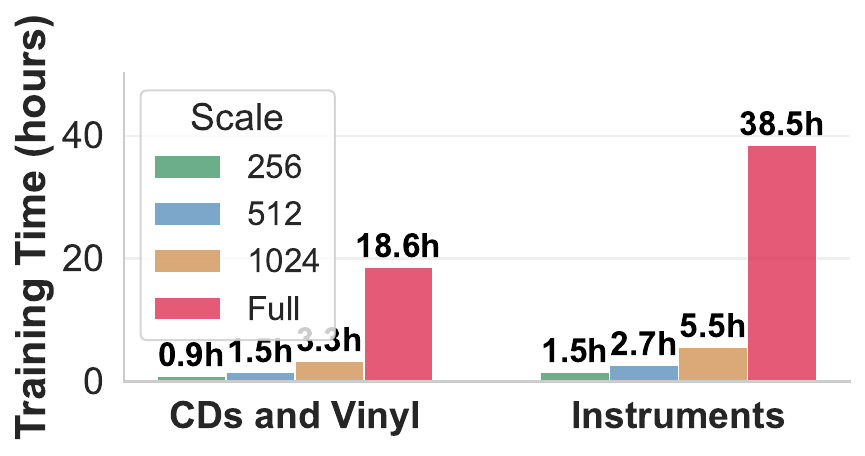}
    % \caption{Training time comparison on the CDs and Vinyl dataset at different data scales (256, 512, 1024, and full).}
    \caption{Training time comparison of MiniRec and full-data fine-tuning on the CDs \& Vinyl and Instruments datasets under different subset sizes ($|\mathcal{S}| \in \{256, 512, 1024\}$). MiniRec achieves substantial time reduction—up to 82\%—while maintaining comparable performance.}
    \label{fig:time}
    \vspace{-15pt}
\end{figure}

This experiment investigates the training efficiency of MiniRec by measuring the reduction in computational cost under different subset sizes. The goal is to evaluate whether MiniRec can accelerate fine-tuning of LLM-based recommender models while retaining high performance. All experiments are conducted on a cluster equipped with four NVIDI A5000 GPUs (24GB memory each). For both datasets—CDs \& Vinyl and Instruments—we vary the subset size $|\mathcal{S}| \in \{256, 512, 1024 \}$ and compare it with full-data training. The total training time (wall-clock hours) is recorded under consistent hyperparameter settings.

As shown in Figure.~\ref{fig:time}, MiniRec substantially lowers training time compared to full-data fine-tuning. On CDs \& Vinyl, training with 1024 samples reduces wall-clock cost from 18.6h to 3.3h, achieving nearly 82\% savings. Similarly, on Instruments, the cost decreases from 38.5h to 5.5h with minimal performance degradation. Smaller subsets further accelerate the process, with proportionate efficiency gains.
In summary, MiniRec enables a favorable cost–performance balance, reducing training time by one order of magnitude without sacrificing model quality. This demonstrates its practicality for efficient LLM-based recommendation in resource-constrained environments.

\begin{figure*}[h]
    \centering
    \includegraphics[width=0.9\textwidth]{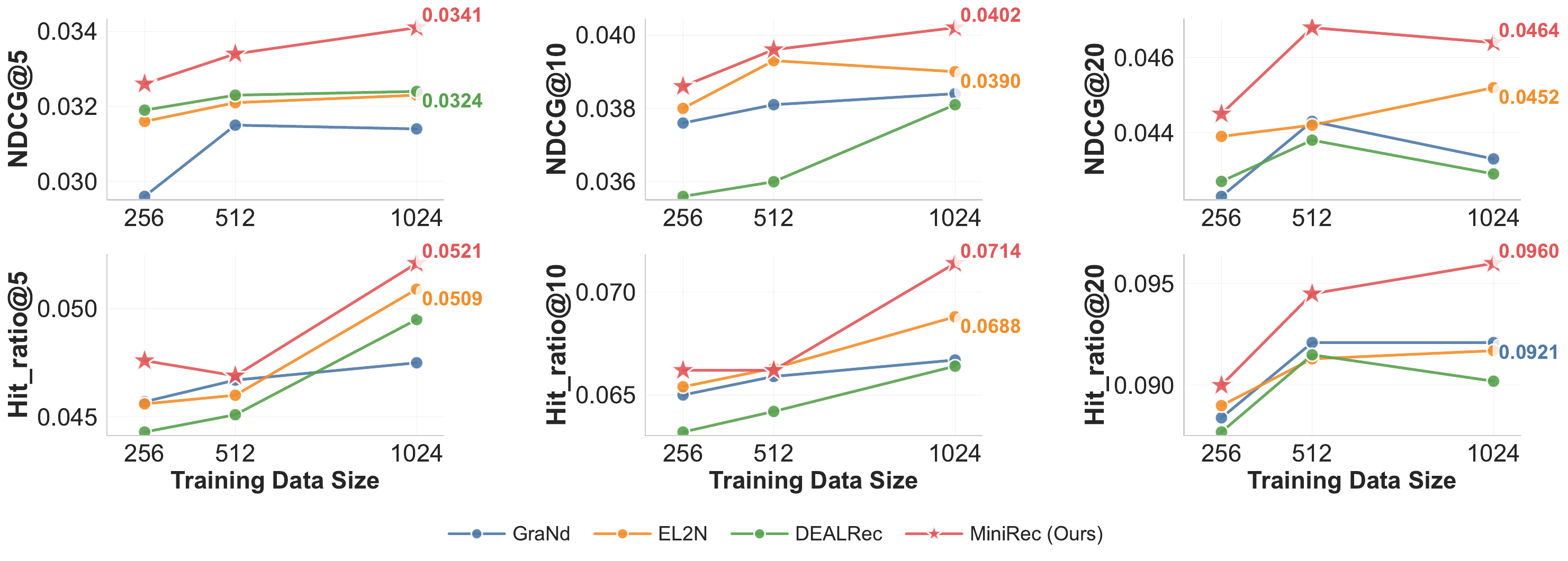}
    \caption{Scalability of MiniRec with varying subset sizes on the CDs \& Vinyl dataset using the \texttt{Gemma‑2‑2B‑it} model. MiniRec consistently outperforms alternative subset selection methods across all sizes and metrics, achieving strong accuracy gains even with small subsets and maintaining efficiency at scale.}
    \label{fig:scale}
\end{figure*}

\subsection{Scalability with Subset Size}
\label{subsec:scalability}
% This experiment evaluates the scalability of MiniRec with respect to subset size and its efficiency–performance trade‑off. Using the Gemma‑2‑2B‑it model on the CDs \& Vinyl dataset, we vary the subset size $|\mathcal{S}| \in \{256, 512, 1024\}$ and assess performance with NDCG@K and HR@K ($K \in \{5, 10, 20\}$).

% As shown in Figure~\ref{fig:scale}, all methods benefit from larger subsets, but MiniRec consistently achieves superior performance at every scale. Even with only 256 samples, it surpasses baselines such as EL2N, and the margin further increases at $|\mathcal{S}|=1024$ (HR@20 = 0.0960). These results indicate that MiniRec selects samples that are both informative and representative, maintaining strong performance under varying data budgets.

% In summary, MiniRec scales effectively with subset size, offering robust accuracy improvements without sacrificing efficiency.

This experiment examines the scalability of MiniRec with respect to subset size, aiming to understand how model performance evolves as more training samples are included. The objective is to evaluate whether MiniRec can maintain its advantage under varying data budgets and to assess its efficiency–performance tradeoff.

We use the Gemma‑2‑2B‑it model on the CDs \& Vinyl dataset, varying the subset size $|\mathcal{S}| \in \{256, 512, 1024\}$. For each scale, the same base LLM and fine-tuning configuration are applied across all subset selection methods to ensure fair comparison. The evaluation uses standard top‑$K$ ranking metrics, including NDCG@K and Hit Ratio@K for $K = \{5, 10, 20\}$.

As reported in Figure~\ref{fig:scale}, all methods exhibit a consistent upward trend in performance as subset size increases, reflecting the natural benefit of larger training sets. However, MiniRec not only achieves the highest metrics at every scale but also demonstrates more pronounced gains than competitors. At $|\mathcal{S}|=256$, MiniRec already outperforms baselines such as EL2N, and the margin widens further at $|\mathcal{S}|=1024$, where it reaches Hit Ratio@20 = 0.0960. These results suggest that samples selected by MiniRec are both informative and representative, effectively capturing diverse recommendation patterns even with limited data.

% In this section, we evaluate MiniRec in terms of effectiveness, generalization, efficiency, and scalability. We present results across multiple datasets and LLM architectures, followed by ablation studies to validate the contribution of each component.
\begin{figure}
    \centering
    \includegraphics[width=0.95\linewidth]{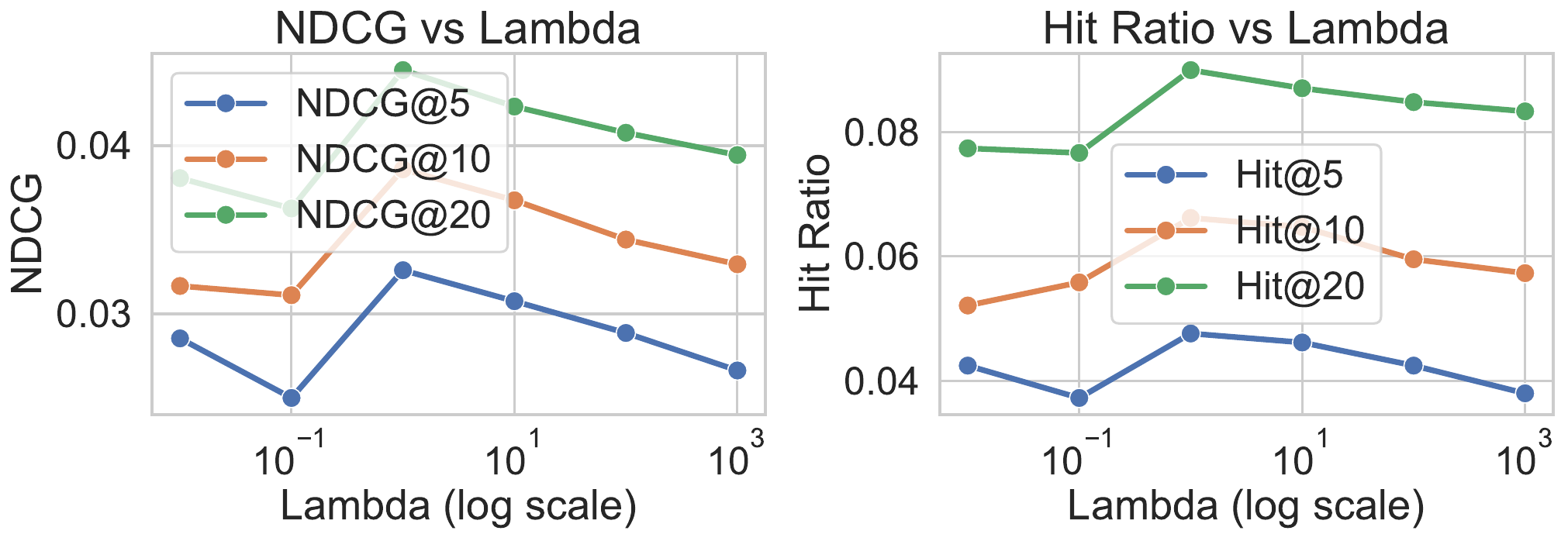}
    \caption{Sensitivity of Gemma‑2‑2B‑it performance to the regularization coefficient $\lambda$ on the CDs \& Vinyl dataset (subset size = 256). Performance improves as $\lambda$ increases from 0.01 to 1, reaching the best results at $\lambda = 1$, after which excessive regularization degrades performance.}
    \label{fig:para}
\end{figure}

\subsection{Parameter Sensitivity Analysis}

% The experiment examines the sensitivity of model performance to the regularization coefficient $\lambda$, which controls parameter regularization strength. Using the Gemma‑2‑2b‑it model with an embedding dimension of 256, $\lambda$ was varied over ${0.01, 0.1, 1, 10, 100, 1000}$, and performance was evaluated with NDCG@K and HR@K metrics ($K \in {5, 10, 20}$).

% As shown in Figure.~\ref{fig:para}, performance improves as $\lambda$ increases from 0.01 to 1, peaking at $\lambda = 1$ (\textit{NDCG@5 = 0.0326}, \textit{HR@5 = 0.0476}). Moderate regularization thus mitigates overfitting and enhances generalization, whereas excessive regularization ($\lambda > 1$) constrains model learning. Consequently, $\lambda = 1$ is adopted as the default setting in subsequent experiments.

This experiment investigates the \textbf{sensitivity} of model performance to the regularization coefficient $\lambda$, which controls the strength of parameter smoothing during optimization. Using the \textit{Gemma‑2‑2B‑it} model with an embedding dimension of 256, we vary $\lambda$ over 0.01, 0.1, 1, 10, 100, 1000
and evaluate performance using standard ranking metrics \textit{NDCG@K} and \textit{Hit Ratio@K} for $K \in \{5, 10, 20\}$.

As shown in Figure~\ref{fig:para}, model performance initially improves as $\lambda$ increases from 0.01 to 1, achieving the best results at $\lambda = 1$ (\textit{NDCG@5 = 0.0324}, \textit{HR@5 = 0.0476}). This indicates that a moderate weighting yields a balanced combination of the scoring terms, leading to stable and effective data valuation. When $\lambda$ becomes excessively large ($\lambda > 1$), one component of the score dominates the others, disrupting this balance and resulting in a performance drop.

Overall, these results highlight the importance of balancing regularization strength in the MiniRec framework. The chosen setting of $\lambda = 1$ offers the best trade‑off between model stability and expressive flexibility, and is therefore adopted as the default configuration in subsequent experiments.

% \vspace{-20pt}
\subsection{Ablation Study}
\label{subsec:ablation}

To quantify the contribution of each design component in MiniRec, 
we conduct an {ablation study} on the {CDs \& Vinyl} dataset using the {Gemma‑2‑2B‑it} model 
with a subset size of $|\mathcal{S}| = 1024$. 
Specifically, we independently remove the following components from the full framework: \emph{Learnability}; \emph{Representativeness}; 
\emph{Diversity}; and \emph{Curriculum scheduling}. 
All other training configurations are kept identical to isolate the effect of each individual module.

As shown in Table~\ref{tab:ablation}, removing any single component leads to noticeable degradation across both 
{NDCG@K} and {Hit Ratio@K} metrics. 
The absence of \emph{Representativeness} or \emph{Diversity} results in clear reductions in ranking quality, 
highlighting their importance in maintaining data coverage and variety. 
Similarly, removing the \emph{Curriculum} mechanism weakens convergence stability and limits the benefit of progressive optimization. 
Among all variants, the complete {MiniRec} configuration achieves the best results 
({NDCG@10} = 0.0402 and {Hit Ratio@10} = 0.0714), 
demonstrating that these components jointly contribute to performance improvement.
The ablation results demonstrate that all modules are complementary, and their joint integration ensures stable, optimal performance in LLM‑based recommendation.

% In summary, the ablation results confirm that each module in MiniRec plays a distinct yet complementary role. 
% Their integration under a unified value function is essential for achieving optimal and stable performance 
% in LLM-based recommendation tasks.

\begin{table}[htbp]
    \caption{Ablation results on the CDs \& Vinyl dataset (\texttt{Gemma‑2‑2B‑it}, $|\mathcal{S}|=1024$). Removing Learnability (L), Representativeness (R), Diversity (D), or Curriculum (C) leads to consistent performance degradation, showing their complementary benefits within MiniRec.}
    \label{tab:ablation}
    \setlength{\tabcolsep}{3pt}
    \begin{tabular}{ccccccc}
    \hline
    \multicolumn{7}{c}{CDs   and Vinyl (1024) Gemma-2-2b-it}                                                                           \\
    \multicolumn{1}{l}{}   & N@5          & N@10         & N@20         & HR@5    & HR@10   & HR@20   \\ \hline
    w/o L       & 0.0315          & 0.0393          & 0.0446          & 0.0476          & 0.0714          & 0.0930          \\
    w/o R & 0.0307          & 0.0373          & 0.0442          & 0.0439          & 0.0647          & 0.0923          \\
    w/o D          & 0.0312          & 0.0385          & 0.0443          & 0.0469          & 0.0692          & 0.0923          \\
    w/o C         & 0.0332          & 0.0400          & 0.0459          & 0.0486          & 0.0699          & 0.0935          \\
    MiniRec                & \textbf{0.0341} & \textbf{0.0402} & \textbf{0.0464} & \textbf{0.0521} & \textbf{0.0714} & \textbf{0.0960} \\ \hline
    \end{tabular}
\end{table}
\section{Conclusion}

In this work, we addressed the challenge of efficient data utilization for reinforcement learning–based large language model (LLM) recommendation. We identified two major limitations of existing data selection approaches—unreliable learnability estimation under reward‑based optimization and misalignment between dataset coverage and representativeness for reasoning‑oriented learning.
To overcome these issues, we proposed MiniRec, a unified data selection framework integrating reward‑driven learnability, gradient‑based representativeness, and diversity‑aware subset construction, enhanced with curriculum scheduling. MiniRec efficiently focuses training on informative and complementary samples, reducing computational cost without compromising performance.
Extensive experiments across multiple real‑world datasets show that MiniRec achieves comparable or superior recommendation quality with significantly less data and training time. Moreover, it generalizes well across LLM architectures and scales effectively with subset size. These findings establish MiniRec as a practical and model‑agnostic solution for data‑efficient reinforcement learning, paving the way toward scalable and reasoning‑aligned LLM‑based recommender systems.

% \begin{acks}
% \end{acks}

% \bibliographystyle{ACM-Reference-Format}
\newpage
\bibliographystyle{ACM-Reference-Format}
\bibliography{sections/0_ref}

\end{document}